\documentclass[useAMS,usenatbib,usegraphicx]{mn2e}
\usepackage{amssymb}
\usepackage{times}
\usepackage{aas_macros} 

\usepackage{epsf}
\usepackage[usenames]{color}

\newcommand{\red}{\textcolor{Red} }

\title[Time resolved spectroscopy of HD~213637]{Time resolved spectroscopy of the cool Ap star HD~213637 \thanks{Based on observations collected at the European Southern Observatory, Chile, as part of programme 077.D-0149(A) }}

\author[Elkin et al.] {V.G.~Elkin$^1$, D.W.~Kurtz$^1$, G. Mathys$^2$ \\
$^1$Jeremiah Horrocks Institute, University of Central Lancashire, Preston PR1~2HE, UK  \\
$^2$Joint ALMA Observatory \& European Southern Observatory, Alonso de Cordova 3107, Vitacura, Santiago, Chile}

\begin{document}

\maketitle

\begin{abstract}
We present an analysis of high time resolution spectra of the chemically peculiar Ap star HD~213637. The star shows rapid radial velocity variations with a period close to the photometric pulsation period. Radial velocity pulsation amplitudes vary significantly for different rare earth elements. The highest pulsation amplitudes belong to lines of Tb~\textsc{iii} ($\sim360$~m~s$^{-1}$), Pr~\textsc{ii} ($\sim250$~m~s$^{-1}$) and Pr~\textsc{iii} ($\sim$230~m~s$^{-1}$). We did not detect any pulsations from spectral lines of Eu~\textsc{ii} and in H$\alpha$, in contrast to many other roAp stars. We also did not find radial velocity pulsations using spectral lines of other chemical elements, including Mg, Si, Ca, Sc, Cr, Fe, Ni, Y and Ba. There are phase shifts between the maxima of pulsation amplitudes of different rare earth elements and ions, which is evidence of an outwardly running magneto-acoustic wave propagating through the upper stellar atmosphere.
\end{abstract}


\section{Introduction}

Rapidly oscillating (roAp) stars are a group of main-sequence cool chemically peculiar A type stars that pulsate in high radial overtone p~modes with periods in the range $6 - 24$~min. They show photometric amplitudes less than 0.01~mag, whereas rapid radial velocity variations in rare earth element lines can reach several km~s$^{-1}$ (e.g., \citealt{Frey09}; \citealt{Kurtz90}). The roAp stars are main-sequence stars with large overabundances of rare earth elements in their upper atmosphere and magnetic fields with strengths ranging from several hundred gauss to several kilogauss. The magnetic field in these stars can described with the oblique rotator model \citep{Stibbs50}. To a first approximation the field may be modelled with a simple dipole structure with the axis of the magnetic dipole inclined to the rotation axis. The roAp stars have effective temperatures ranging between 6400 and 8200~K. 

The instability strip crosses the main-sequence in this temperature region where variables of $\delta$~Sct type are found. Pulsation in $\delta$~Sct stars is driven by the $\kappa$-mechanism in the He~\textsc{ii} ionization zone, whereas pulsation in roAp stars is excited in the hydrogen partial ionization zone by the $\kappa$-mechanism near the magnetic poles where convection is suppressed (\citealt{Balmforth01}; \citealt{Saio05}; \citealt{Cunha02}). The relationship between pulsation and other properties of roAp stars, such as their strong magnetic fields and chemical anomalies, is still not completely clear, hence requires more observational and theoretical research.

Rapid oscillations were first found in Ap stars by \citet{Kurtz78} in one of the coolest peculiar stars, HD~101065. A number of observing programmes using different telescopes and methods have now discovered about 60 roAp stars. Among them is one of the coolest chemically peculiar stars, HD~213637, for which \citet{Martinez98} found two main photometric frequencies, $1452.35 \pm 0.06~\mu$Hz ($P = 11.5$~min) and $1410.89 \pm 0.11~\mu$Hz ($P = 11.8$~min). The photometric pulsation amplitudes are less than 1.5~mmag in Johnson $B$ and vary from night to night; they are nearly undetectable on some nights. These amplitude variations may be a result of rotational modulation or beating between different mode frequencies. Probably for these reasons and the rather small amplitude \citet{Nelson93} failed to detect the pulsations in HD~213637.

\begin{figure}
\begin{center}
\hfil \epsfxsize 8.5cm\epsfbox{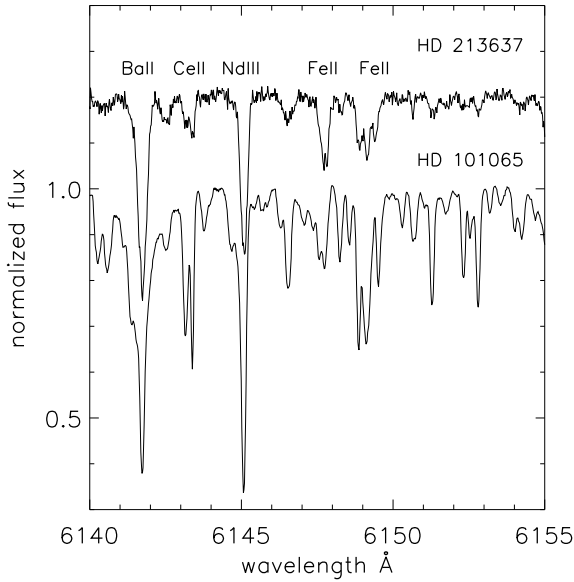}
\caption{\label{fig:spec1} A portion of the spectrum of HD~213637 compared to one of the most peculiar stars, HD~101065. The spectrum of HD~213637 is shifted on the y-axis for convenience. Several strong spectral lines are identified. Both stars show strong lines of rare earth elements.}
\end{center}
\end{figure}

The main-sequence position of HD~213637 is especially interesting. This star is at the low temperature border of the roAp stars, as well as all other Ap stars, making it one of the coolest Ap stars known. Its effective temperature of $T_{\rm  eff} = 6400$~K is close to that of the first detected roAp star, HD~101065, although the peculiarity of HD~213637 is not as extreme as that of HD~101065. Fig.~\ref{fig:spec1} compares a portion of the spectra of HD~213637 and HD~101065. Both spectra have strong lines of rare earth elements. Both stars lie at the low temperature border of magnetic chemically peculiar stars including roAp stars, but HD~101065 has been more extensively discussed in the literature.

While initially roAp stars were discovered and studied with photometric methods, time resolved spectroscopy has allowed the study of wider physical aspects of the pulsating stellar atmosphere. Spectral lines of chemical elements form in different atmospheric layers in Ap stars; the atmospheres are stratified by atomic diffusion. The rapid radial velocity variations of spectral lines of certain chemical elements allow us to sample the velocity field in the stellar atmosphere as a function of atmospheric depth. In this paper we present the results of a high time resolution spectroscopic study of HD~213637 and discuss the pulsation behaviour of different chemical elements. 

\section{Observations and data reduction}

High-resolution spectra of HD~213637 were obtained with the Ultraviolet and Visual Echelle Spectrograph (UVES) on the 8-m Very Large Telescope (VLT) of the European Southern Observatory (ESO). In total 75 spectra were obtained with exposure times of 80~s and readout plus overhead times about of $\sim$21~s, corresponding to a time resolution of $\sim$101~s. A 0.3~arcsec slit and image slicer were used. The echelle spectra cover the wavelength range $\lambda\lambda~4970 - 7010$~\AA, with a 60--70~\AA\ gap around 6000~\AA\ caused by the space between the two CCDs. The average spectral resolution is about $R = 10^5$. The CCD frames were processed using {\small ESO-MIDAS} and the UVES pipeline to extract and merge the echelle orders to 1D spectra that were normalized to the continuum.

\subsection{The effective temperature}

The atmospheric parameters and chemical abundances in HD~213637 were discussed by \citet{Koch03}. From Str\"omgren photometry by \citet{Martinez98} he found an effective temperature of $T_{\rm eff} = 6700$~K and $\log g = 4.6$ (cgs units). By fitting the Balmer H$\alpha$ and H$\beta$ lines with synthetic profiles \citet{Koch03} determined an effective temperature of $T_{\rm eff} = 6400 \pm 100$~K. From the ionization equilibrium of Fe~\textsc{ii} and Fe~\textsc{i} he estimated a surface gravity $\log g = 3.6 \pm 0.2$, which is smaller than the result from photometry. 

We used Str\"omgren photometry \citep{Martinez93} and the calibration by \citet{Moon85}, which give for HD~213637 an effective temperature of $T_{\rm eff} = 6400$~K. We calculated an average spectrum from the first 55 UVES spectra. This spectrum was used for identification of spectral lines and for determination of stellar parameters. We calculated a synthetic spectrum of the star using the {\small SYNTH} code of \citet{piskunov92}. For synthetic calculations and line identification we used a spectral line list extracted from the Vienna Atomic Line Database (VALD \citealt{kupka99}), which includes lines of rare earth elements from the DREAM database \citep{biemontetal99}. Atmosphere models were taken from the NEMO (Vienna New Model Grid of Stellar Atmospheres) database \citep{heiter02}.  From fitting the H$\alpha$ profile of an average spectrum of HD~213637 with a synthetic profile we determined $T_{\rm eff} = 6400 \pm 200 $~K, which is in agreement with \citet{Koch03} and with the photometric results. This effective temperature is one of the lowest among all chemically peculiar magnetic stars. 

\begin{figure}
\resizebox{\hsize}{!}{\includegraphics[angle=270]{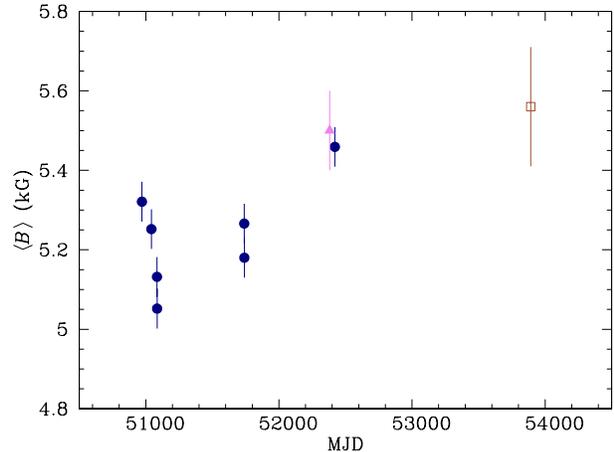}}
\caption{Mean magnetic field modulus of the star HD~213637,
against Modified Julian Date: measurements of \citet{Mathys03} (dots), \citet{Koch03} (triangle), and this paper (open square).}
\label{fig:hm213637}
\end{figure}

\subsection{The magnetic field}

A strong global magnetic field with a predominantly dipolar structure is one of the characteristics of roAp stars. This magnetic field can be detected in high-resolution spectra using magnetic broadening or resolved Zeeman components of spectral lines \citep{Mathys97}. The spectropolarimetric technique allows the measurement of the longitudinal component of the average magnetic field vector. 

HD~213637 has a strong global magnetic field. \citet{Mathys03} detected magnetically split lines in its spectrum and reported an average magnetic field modulus of 5.2~kG. This average was computed by combining measurements (to be published elsewhere) of the split components of the line Fe~\textsc{ii}~$\lambda\,6149$~\AA\ in spectra taken at seven different epochs over a timespan of $\sim$4~y, yielding field modulus values ranging from 5.05 to 5.45~kG, whose spread definitely reflects actual variations. \citet{Koch03} derived a value of $5.5 \pm 0.1$~kG for the magnetic field modulus, from consideration of the same line. This value does not significantly differ from that obtained from the final observation of Mathys, 39~d later. It is also close to the value of $5.56\pm0.15$~kG that we determined using our average UVES spectrum, recorded $\sim$4~y after Mathys's final observation. As the split components of  Fe~\textsc{ii}~$\lambda\,6149$~\AA\ are only partially resolved, and as the line is affected by a blend on the blue side (as in many other Ap stars with resolved magnetically split lines -- see Figs~2 to 4 of \citealt{Mathys97}), we used Gaussian fitting to get the central positions of the resolved parts of both components. 

All the measurements of the magnetic field modulus of HD~213637 described above are plotted against the Modified Julian Date in Fig.~\ref{fig:hm213637}. The distribution of the observing points in time does not lend itself to unambiguous determination of the period of variation of the magnetic field (that is, the rotation period of the star). However, this period appears unlikely to be shorter than $\sim10$ days, and it may possibly be several months.

In addition, some other spectral lines with large Land\'e factors also show magnetic broadening and partial Zeeman splitting. \citet{Mathys03} also obtained one observation of the longitudinal magnetic field of 0.2~kG. \citet{Hubrig04}, using spectropolarimetric observations with the FORS1 (FOcal Reducer and low dispersion Spectrograph) instrument on the VLT, found a longitudinal component of the magnetic field of $740 \pm 50$~G. The difference between the longitudinal magnetic field measurements of \citet{Hubrig04} and \citet{Mathys03}, which were obtained more than 4.5~y apart, is probably the result of observations taken at different phases of the unknown rotation period for this star.

\section{Radial velocity variations}

In roAp stars spectral lines of rare earth elements normally show rapid radial velocity variations with the same periods found in the photometric pulsations (\citealt{Malanushenko98}; \citealt{Koch01}; \citealt{Kurtz07}). The lines of other chemical species, including light elements and iron peak elements, show much smaller pulsation amplitude, or show none at all. This behaviour can be interpreted as an effect of stratification in the stellar atmosphere. The rare earth elements concentrate higher in the atmosphere at optical depth above $\log\tau_{5000}=-3.0$ \citep{Mash05} where oscillation amplitudes reach a maximum, while light and iron peak elements tend to concentrate in the bottom of the atmosphere at optical depth $\log\tau_{5000}=-1.0$ where the pulsation amplitude is lower \citep{Ryab02}. In this situation we have a spatial filter that allows the study of the velocity field in different layers in the stellar atmosphere by using spectral lines of different elements and ions. To explore the velocity fields in the atmospheres of roAp stars we need to detect the pulsational Doppler shift of individual spectral lines at different phases of the pulsation period. It is necessary to identify the spectral lines carefully, which we did using synthetic spectra of the star calculated with the {\small SYNTH} code of \citet{piskunov92}. 

For Doppler shift measurements we calculated the central positions for profiles of individual spectral lines by the centre-of-gravity method for each spectrum in the time series. Frequency analysis of radial velocities, and determination of the amplitudes and phases were performed with least-squares fitting using {\small ESO-MIDAS}'s Time Series Analysis, a discrete Fourier transform programme by \citet{kurtz85} and the Period04 software by \citet{lenz05}. We have just 2~h of observations, which are not sufficient to resolve the two main photometric frequencies, 1452.35 and 1410.89~$\mu$Hz, found by \citet{Martinez98}. From lines with high pulsation amplitudes we found an average pulsation frequency of 1434~$\mu$Hz that was used for calculating amplitudes and phases by least-squares fitting for individual spectral lines and for combination of lines. These combinations of lines were used to improve precision and better study the pulsation characteristics.

\subsection{Lines of rare earth elements}

\begin{figure}
\begin{center}
\hfil \epsfxsize 8.5cm\epsfbox{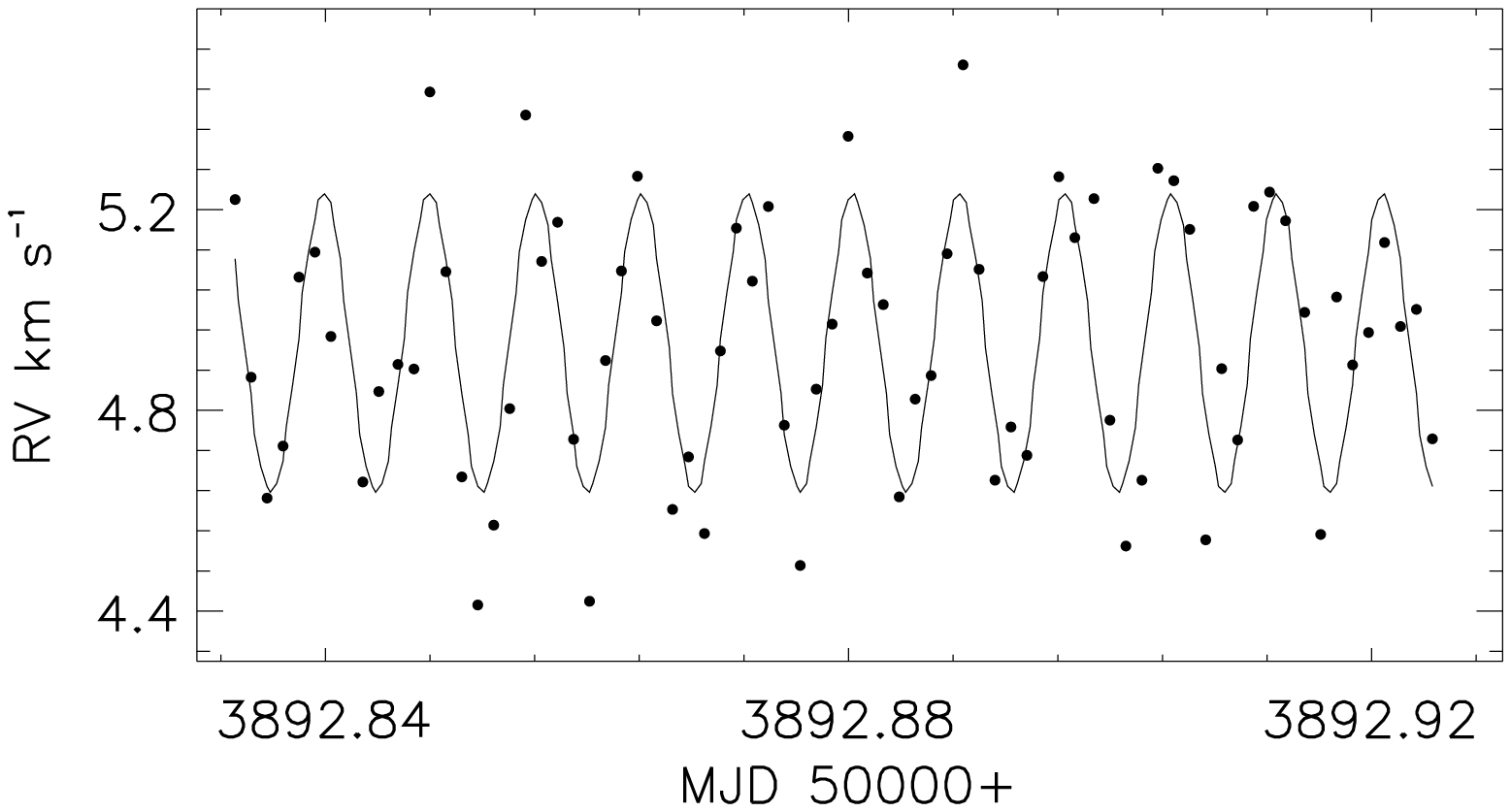}
\hfil \epsfxsize 8.5cm\epsfbox{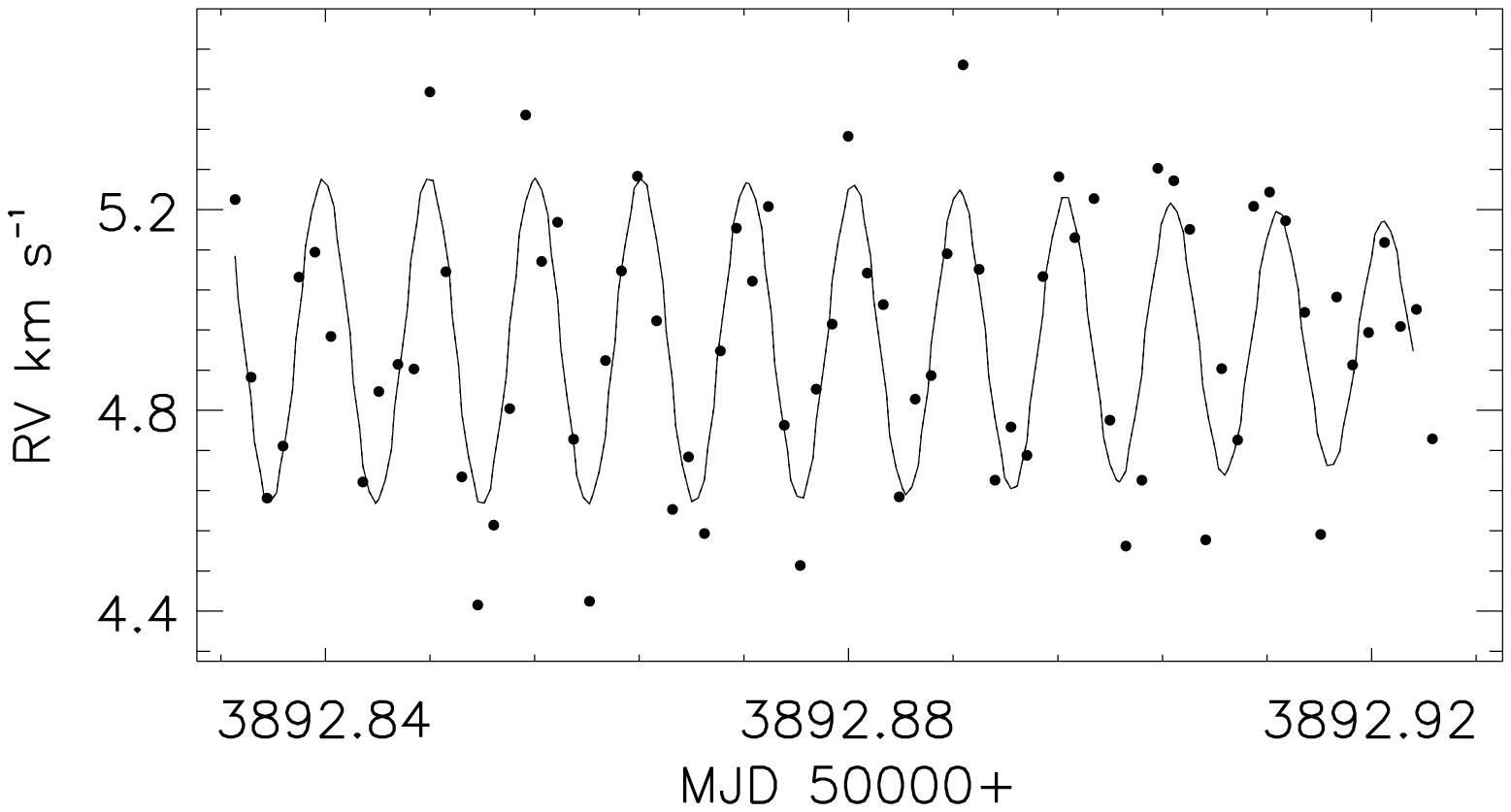}
\caption{\label{fig:rvcurv} Top: The radial velocity curve measured from the Tb~\textsc{iii} $\lambda\,5505$~\AA\ line. The solid curve is a least-squares fit using the average frequency of 1434~$\mu$Hz. Bottom: This is the same radial velocity curve fitted with the two photometric frequencies of $1411~\mu$Hz and $1452~\mu$Hz.}
\end{center}
\end{figure}

We identified spectral lines of HD~213637, many of which belong to rare earth elements. In our spectra the most numerous lines belong to ions of neodymium and cerium. The pulsation amplitudes of spectral lines in HD~213637 vary for different elements and ions. The highest amplitudes, up to 360~m~s$^{-1}$, were found in lines of Tb~\textsc{iii}. The top panel of Fig.~\ref{fig:rvcurv} shows the radial velocity curve for the Tb~\textsc{iii} $\lambda\,5505$~\AA\ line. The solid curve is a least-squares fit for the average frequency 1434~$\mu$Hz. The bottom panel shows the same radial velocity curve fitted with the two photometric frequencies. There is little difference between these curves, as the amplitude of the second frequency is small. 

\begin{figure}
\begin{center}
\hfil \epsfxsize 8.5cm\epsfbox{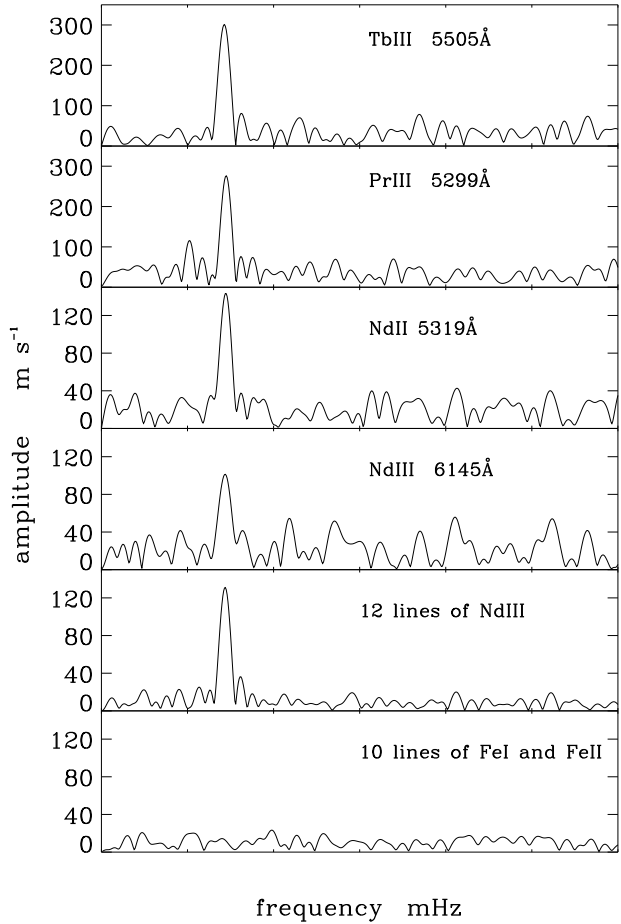}
\caption{\label{fig:ampspec1} Amplitude spectra for HD~213637 for individual lines and for combinations of lines of the same chemical element. Note the difference in the ordinate scale between the top two and bottom four panels.}
\end{center}
\end{figure}

Examples of amplitude spectra for individual lines of Tb~\textsc{iii} $\lambda\,5505$~\AA, Pr~\textsc{iii} $\lambda\,5299$~\AA, Nd~\textsc{ii} $\lambda\,5319$~\AA\ and Nd~\textsc{iii} $\lambda\,6145$~\AA\ are shown in the top four panels in Fig.~\ref{fig:ampspec1}. In this figure the two bottom panels illustrate amplitude spectra for combinations of twelve lines of Nd~\textsc{iii} and ten lines of Fe~\textsc{i} and Fe~\textsc{ii}.

A summary of radial velocity amplitudes and phases is given in Table~\ref{tab:RV_res}, calculated from a combination of radial velocities obtained from $1 - 25$ lines for each ion using an average frequency of 1434~$\mu$Hz and a time zero point of MJD~53892.050. The differences in the amplitudes and phases for different chemical elements and ions reflect where their spectral lines form in the stellar atmosphere. The highest average amplitude is seen for lines of Tb~\textsc{iii} and these lines also show a phase shift compared with other rare earth element lines. There are other roAp stars, such as HD~176232, that also have the highest pulsation amplitude detected from lines of Tb~\textsc{iii} among the other rare earth element lines \citep{Elkin08b}. For several other roAp stars with narrow lines \citet{Ryab07b} found that lines of Tb~\textsc{iii} have the highest pulsation amplitude and largest phase shift. This is probably the result of the Tb~\textsc{iii} line-forming layer being the highest in the atmosphere of all the rare earth elements studied. The lines of Pr~\textsc{ii} and Pr~\textsc{iii} have an average pulsation amplitude of $\sim$240~m~s$^{-1}$. These lines also have similar phases, which means that they form in the same atmospheric layer. The lines of La~\textsc{ii}, Ce~\textsc{ii}, Nd~\textsc{ii}, Nd~\textsc{iii} and Sm~\textsc{ii} show pulsation amplitudes from 45 to 140~m~s$^{-1}$. 

\begin{table} 
\caption[]{Pulsation amplitudes and phases from radial velocity measurements for various ions in HD~213637. The phases were calculated for the frequency 1434~$\mu$Hz and a time zero point of MJD~53892.050. No phase is given for ions for which the amplitude is not significant. \red{The ions are listed in order of increasing pulsation amplitude.}} 
\label{tab:RV_res} 
\begin{center}  
\begin{tabular}{lrrc} 
\hline  
\multicolumn{1}{c}{Ion} &  
\multicolumn{1}{c}{Number} &  
\multicolumn{1}{c}{Amplitude} & 
\multicolumn{1}{c}{Phase} \\ 
 & \multicolumn{1}{c} {of lines}  & 
\multicolumn{1}{c}{m~s$^{-1}$} &  
\multicolumn{1}{c}{Rad} \\ 
\hline  
Y~\textsc{ii}   &   4  &  $  4 \pm   12 $ & $                  $  \\ 
Fe~\textsc{i} \& Fe~\textsc{ii} &  10  &  $  13 \pm   \phantom{0}8 $ & $               $  \\ 
H$\alpha$ core    &   1  &  $  26 \pm   20 $ & $                  $  \\  
Eu~\textsc{ii}   &   3  &  $  27 \pm   36 $ & $                  $  \\ 
La~\textsc{ii}   &   9  &  $  44 \pm   13 $ & $ 1.866 \pm  0.301 $  \\ 
Ce~\textsc{ii}   &  12  &  $  90 \pm   14 $ & $ 1.430 \pm  0.151 $  \\ 
Sm~\textsc{ii}   &   6  &  $  109 \pm   14 $ & $ 1.419 \pm  0.131 $  \\ 
Nd~\textsc{iii}  &  12  &  $  131 \pm   \phantom{0}7 $ & $ 1.549 \pm  0.050 $  \\ 
Nd~\textsc{ii}   &  25  &  $  139 \pm   \phantom{0}6 $ & $ 1.551 \pm  0.045 $  \\ 
Pr~\textsc{iii}  &   8  &  $  233 \pm   \phantom{0}9 $ & $ 1.390 \pm  0.038 $  \\ 
Pr~\textsc{ii}   &   5  &  $  253 \pm   21 $ & $ 1.401 \pm  0.085 $  \\ 
Tb~\textsc{iii}  &   4  &  $  356 \pm   21 $ & $ 0.908 \pm  0.058 $  \\ 

\hline  
\end{tabular}                
\end{center} 
\end{table}

\begin{figure}
\begin{center}
\hfil \epsfxsize 8.5cm\epsfbox{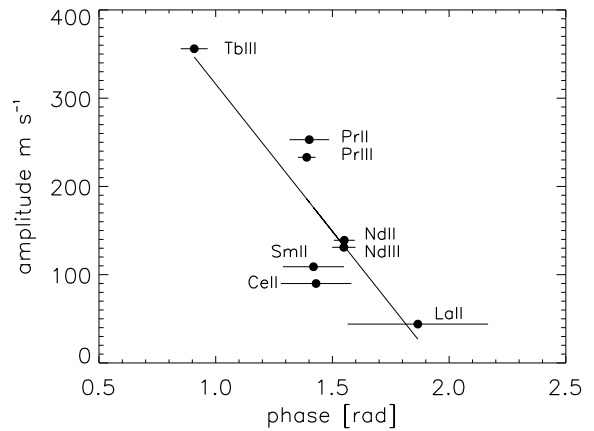}
\caption{\label{fig:amph1} Average amplitudes and phases for lines of rare earth elements. The solid line is a least-squares linear fit. }
\end{center}
\end{figure}

There are phase shifts between various rare earth element lines that are illustrated in Fig.~\ref{fig:amph1}. The correlation of the amplitudes and phase shifts is explained by outwardly propagating magneto-acoustic waves. Conservation of pulsational kinetic energy leads to an increase in amplitude higher in the atmosphere where the density is lower. The phases are defined such that larger phases indicate earlier times of pulsation maxima, hence the plot shows the outwardly running character of the pulsations, from the deeper La~\textsc{ii} forming layer through the region of formation of other rare earth elements, including Ce~\textsc{ii}, Nd~\textsc{ii}, Nd~\textsc{iii} and Sm~\textsc{ii} and  finally to the highest observable, lowest density layers where the Pr~\textsc{ii}, Pr~\textsc{iii} and Tb~\textsc{iii} lines form. 

In HD~213637 we found no significant radial velocity pulsations from lines of Eu~\textsc{ii}. This is in contrast with some other roAp stars. Pulsation amplitudes obtained from Eu~\textsc{ii} lines vary from star to star. For example, HD~99563 shows the strongest pulsation amplitudes for Eu~\textsc{ii} lines \citep{Elkin05b}. Significant pulsations from Eu~\textsc{ii} were detected in HD~166473 \citep{Mathys07} and in HD~176232 \citep{Elkin08b}. But \citet {Ryab07b} mentioned that Eu~\textsc{ii} lines show the lowest pulsation amplitudes for rare earth element lines in several roAp stars. Why should the pulsation amplitudes from Eu~\textsc{ii} lines be so different in various roAp stars?

For the roAp star HR~3831 \citep{Koch04} calculated distribution maps of chemical elements over the stellar surface using a Doppler imaging technique \citep{Pisk93}. This map shows that europium forms in two small spots, where one is shifted from the positive magnetic pole, and the other is between the negative magnetic pole and the magnetic equator.  In HD~99563 \citet{Frey09}, using high resolution spectra from the VLT and Subaru telescopes, calculated a Doppler imaging map for Eu~\textsc{ii} that shows several small spots with high overabundances of europium. Thus it appears that Eu~\textsc{ii} has widely variable surface distributions in roAp stars, which leads to a range of observed amplitude ratios with respect to other rare earth elements. The pulsations in roAp stars are to first order typically dipole modes roughly aligned with the magnetic fields, hence abundance spots near to the pulsation poles give rise to higher amplitudes than spots far from the poles. In our spectra of HD~213637 the lines of Eu~\textsc{ii} are not among the strongest rare earth elements, and are much weaker, for example, than lines of Nd~\textsc{iii}. Most probably the Eu~\textsc{ii} lines we measured form in a small spot, or spots, at some distance from the magnetic poles. 

\subsection{Radial velocity of the H$\alpha$ core}

The roAp stars show a narrow core in the Balmer lines, particularly in the H$\alpha$ profile (\citealt{Wegner76}; \citealt{Cowley01}). \citet{Cowley01} call this the `core-wing anomaly', and show it for a set of cool Ap stars with sharp transitions between wide Stark wings and narrower line cores. \citet{Koch02} suggested that this anomaly results from a hot layer in the abnormal atmosphere at optical depth $\log\tau_{5000}$ between $-2.5$ and $-1.5$. The temperature in this hot layer is higher than in normal stars by about 500~K. For Ap stars with effective temperatures less than 7000~K the proposed hot layer extends to $\log\tau_{5000}\sim -3.5$ with a temperature increase of about 1000~K.

We measured the radial velocity of the narrow core of H$\alpha$ for HD~213637. This part of the profile forms in the upper layers of the stellar atmosphere, thus measures the velocity field at these levels. The wide wings of H$\alpha$ do not allow the radial velocity for the whole line to be measured with suitable precision, but the core of this Balmer line is very narrow and gives the opportunity to detect pulsations. In addition, the core is formed in the atmospheric layers where the pulsation amplitudes increase. The core of the H$\alpha$ profile is thus a good indicator of radial velocity pulsations in roAp stars. Most of the 27 roAp stars studied by \citet{Elkin08a} show pulsation variability in H$\alpha$ core. The radial velocity measurements in the H$\alpha$ core for HD~213637 are shown in Table~\ref{tab:RV_res}. We did not detect significant pulsations, which is unusual for an roAp star. 

The distribution of hydrogen is more uniform than iron peak and rare earth elements. Note, however, that even hydrogen is not uniform over the stellar surface, as shown in the high amplitude roAp star HD~99563 by \citet{Frey09}. This is a result of extreme overabundances of rare earth elements near the magnetic poles in that star. In general, the radial velocities of lines of hydrogen reflect an average value over stellar surface, while measurements from lines of elements concentrated in spots reflect the velocities in the region of the spot only. The pulsation amplitude obtained from lines that form in spots close to the magnetic poles is higher than the average over visible stellar surface obtained from measurements of the Balmer lines cores, as expected for basically dipolar pulsations approximately aligned with the magnetic field. This is partially supported by recent observations of magnetic fields by \citet{Kudr12} who measured longitudinal magnetic fields in 23 chemically peculiar  stars from the cores of Balmer lines and from metal lines. They found that for 22 stars the magnetic field measured from the Balmer line core is smaller than that measured from metal lines. For example, for the known roAp star HD~201601 ($\gamma$~Equ) they determined an average longitudinal magnetic field of $-$1140~G for the metal lines and -390~G for the core of H$\beta$.  \citet{Kudr12} suggested that this difference is evidence of a magnetic field gradient in the stellar atmosphere.

\subsection{Other metal lines without rapid oscillations}

We measured other spectral lines belonging to different chemical elements, including those of Mg, Si, Ca, Sc, Cr, Fe, Ni, Y and Ba. None of these showed any significant pulsations. The bottom panel of Fig.~\ref{fig:ampspec1} shows an amplitude spectrum for a combination of ten Fe~\textsc{i} and Fe~\textsc{ii} lines with no detectable pulsation. This is in agreement with many other roAp stars. These elements have line forming layers deeper in the atmosphere than the rare earth elements, hence show lower pulsation amplitudes. In only a few larger amplitude roAp stars are low amplitude pulsations detected in these metal lines.  In contrast with HD~213637,  in the other very cool roAp star HD~101065 pulsations with amplitudes at the level of several dozens of m~s$^{-1}$ were found for lines of Ca, Ti, Cr, Co, Y, Zr and Ba (Elkin et al., in preparation).  

\section{Discussion and conclusion}

HD~213637 is important to the study of roAp stars, since with an effective temperature $T_{\rm eff} = 6400 \pm 200$~K  it is one of the coolest such stars known, at the lower boundary of the region occupied by this type of stars on the main-sequence. The other example of such a cool roAp star is HD~101065, also with an effective temperature $T_{\rm eff} = 6400$~K (\citealt{Shulyak10}),  which is arguably the most peculiar Ap star known (some might even say the most peculiar star known) and which shows higher pulsation radial velocity amplitudes. For example, HD~101065 has amplitudes which reach almost 0.9~km~s$^{-1}$ for lines of Pr~\textsc{iii} and up to about 0.2~km~s$^{-1}$ in Ce~\textsc{ii} lines (Elkin et al. in preparation). It also has one of the highest photometric pulsation amplitudes of any of the roAp stars (\citealt{Kurtz06}), and was the first one discovered (\citealt{Kurtz78}). 

There are few other roAp stars with effective temperatures below 7000~K. One of them is HD~143487 \citep{Elkin10} with $T_{\rm eff} = 6400$~K; it also shows low amplitude pulsations, so that only the combination of several lines show significant peaks in the amplitude spectra. For this star the average pulsation amplitude shown by lines of Ce~\textsc{ii} is around 70~m~s$^{-1}$; it is even lower, close to 30~m~s$^{-1}$, for lines of Nd~\textsc{iii};  no pulsations were detected for the H$\alpha$ core \citep{Elkin10}. Consideration of HD~101065, HD~143487 and HD~213637 demonstrates that for the coolest roAp stars radial velocity pulsation amplitudes range from hundreds of m~s$^{-1}$ down to only tens of m~s$^{-1}$.    

Pulsations in roAp stars are complex low-degree magneto-acoustic dipole modes, which are distorted by the magnetic field. There is significant magnetic pressure in the stellar atmosphere, which is comparable to, or even higher than the gas pressure \citep{Saio05}. The chemical elements and ions have inhomogeneous distributions in both the horizontal and vertical directions in the atmospheres of roAp stars (\citealt{Koch06}; \citealt{Frey09}). When comparing a theoretical instability strip in the HR~Diagram (e.g., \citealt{Cunha02}; \citealt{Theado09}) with observations of roAp stars, there are discrepancies at both the cooler  and hotter boundaries. The disagreement is more obvious at the cooler edge of the instability strip, where there are roAp stars such as HD~213637 that are cooler than the theoretical boundary. Therefore, this cool roAp star, whose study is presented here, is of importance in understanding the driving and visibility of pulsations in roAp stars.

\section{Acknowledgments}

 This research  made use of NASA's Astrophysics Data System and SIMBAD database, operated at CDS, Strasbourg, France.

\bibliography{arXiv_213637}

\end{document}